\documentclass[seceq]{ptptex}

\usepackage[dvips]{epsfig}

\usepackage{graphicx}
\usepackage{epsfig}

\def\half{\textstyle{\frac{1}{2}}}
\def\quarter{\textstyle{\frac{1}{4}}}

\title{
The three burials of Melquiades DGP%
}

\author{
Ruth \textsc{Gregory}%
}

\inst{
Centre for Particle Theory, 
Durham University, South Road, Durham, DH1 3LE, UK
}

\abst{
In this talk I review three fatal flaws of the DGP braneworld model, 
which has been put forward as a possible model for late time 
acceleration without a cosmological constant; Ghosts, Cosmological
Crashes, and Instability of the 5D vacuum. The talk is based on work in 
collaboration with Charmousis, Kaloper, Myers and Padilla\cite{CGKP,GKMP}.
(The title refers to a film: The three burials of Melquiades Estrada).
}

\begin{document}

\maketitle

\section{Overview}

Current observations\cite{snova} tell us that the 
universe is gently accelerating,
with an effective cosmological constant of about $10^{-30}$g cm$^{-3}$,
yet it is difficult within the context of field theory to get a natural
and consistent explanation of such a value, which is 120 orders of
magnitude smaller than would naively be expected.
On the other hand, Einstein gravity is experimentally verified 
from about 0.1mm to several Kpc. Thus, a natural alternative to 
modifying the matter content of the Universe is to alter the gravitational
interaction at large scales. One particularly attractive 
framework in which to achieve this goal is large extra
dimensions and braneworlds\cite{LED,RS}.

Braneworlds are slices through spacetime on which we live. 
Standard model physics is confined to the brane, but gravity samples 
all of the dimensions. 
The extra dimensions can be strongly warped, or flat, and
the brane itself can have various terms in its effective action:
\begin{equation}
S=-2M_5^3\int_\textrm{bulk} \sqrt{-g} \left ( R+2\Lambda \right )
+ \int_\textrm{brane} \sqrt{-\gamma} \left [ 
- 4M_5^3 K - M_4^2 \mathcal{R} - \sigma + \mathcal{L}_{matter}\right]
\label{action}
\end{equation}
where $\Lambda$ is a possible bulk cosmological constant, $\sigma$
is the brane tension (branes typically have an energy momentum
proportional to the induced metric), $K$ is the extrinsic curvature
of the brane, which is the Gibbons-Hawking term present when we regard
the brane as a boundary to the bulk, and $\mathcal{R}$ is the intrinsic
Ricci curvature of the brane, and represents the DGP\cite{DGP} term.

All of these components influence the effective gravitational 
interaction on the brane. For example, the Randall-Sundrum (RS) 
model \cite{RS} has a positive tension brane living in anti-de
Sitter (adS) spacetime with a fine tuning relation $\sigma^2 = -24
M_5^6 \Lambda$, and has geometry:
\begin{equation}
ds^2 = e^{-2k|z|} \left [ \eta_{\mu\nu} dx^\mu dx^\nu \right] -dz^2
\end{equation}
Geometry away from the braneworld (at $z=0$) is strongly {\it warped},
and the spacetime is symmetric around the brane. Gravity is 4D Einstein
gravity with short range power law corrections.
DGP branes are a different type of brane. They live in Minkowski spacetime
and do not necessarily carry tension. They have the status of a 
brane because of the induced Ricci scalar term. More general
DGP models include a tension for the brane, and possibly a bulk cosmological
constant like the RS model. Four-dimensional gravity comes from from
the imposed brane Einstein-Hilbert term, which dominates at large momenta,
but at low momenta the extra dimension opens up, as in the GRS
model \cite{GRS}, and gravity becomes five-dimensional in nature.

To get cosmological braneworld solutions, we recall that in standard 4D 
cosmology, considerable simplification results from taking the
geometry to be homogeneous and isotropic. In the case of the braneworld, 
making the same assumption results in a slightly more complex
system, as the metric now depends on time and the bulk
distance, however, the Einstein equations are integrable\cite{BCG}, and the
system reduces to a known bulk spacetime (Schwarzschild), with a 
single variable representing the location of the brane in this bulk. 
Using the equations of motion from (\ref{action}) yields the equivalent
of the Friedmann equation:
\begin{equation}
\label{frweq}
\epsilon \left [ \frac{\kappa}{R^2} - \frac{\mu}{R^4} - \frac{\Lambda}{6}
+ H^2 \right ] ^{1/2} = \frac{E}{12M_5^3} - \frac{M_4^2}{2M_5^3}
\left ( H^2 + \frac{\kappa}{R^2} \right )
\end{equation}
In this equation, $\mu$ is a mass parameter for the bulk black 
hole, $\kappa$ is the curvature of the Universe, and $E = \sigma + \rho$ 
is the total energy density (brane plus matter density) of the 
Universe. The parameter $\epsilon=\pm1$ is the sign of the outward 
pointing normal of the brane, and tells us which side of the brane is the
bulk spacetime. Note that in the absence of the DGP term, the sign
of $\epsilon$ is linked to the sign of the energy density of the brane.
For example, the pure RS brane has $H=\kappa=\mu=0$,
giving the fine tuning condition as stated. RS brane cosmology has
again no $M_4$-term, and inputting $E= 12M_5^3 k + \rho$ gives
the well-known non-conventional cosmology equations\cite{NCC}.

Finding cosmological solutions in the DGP model is straightforward, as
the integrability argument is dependent on the symmetry rather than
the specific model, thus the bulk spacetime is either flat or Schwarzschild. 
The effect of the DGP term as can be seen from (\ref{frweq})
is to add a geometric term to the brane energy, thus allowing for
both signs of $\epsilon$ while maintaining positive brane energy. Setting
$\mu=\Lambda=\kappa=E=0$ to get the original tension-free DGP brane, gives 
two solutions: $\epsilon=1, H=0$, the {\it normal
branch} (N), the original solution of DGP\cite{DGP}, and 
$\epsilon=-1,\ H=2M_5^3/M_4^2$, a pure de-Sitter universe: the
{\it self-accelerating} (SA) branch\cite{DGPcos}.

\begin{figure}
{\centerline{\epsfig{file=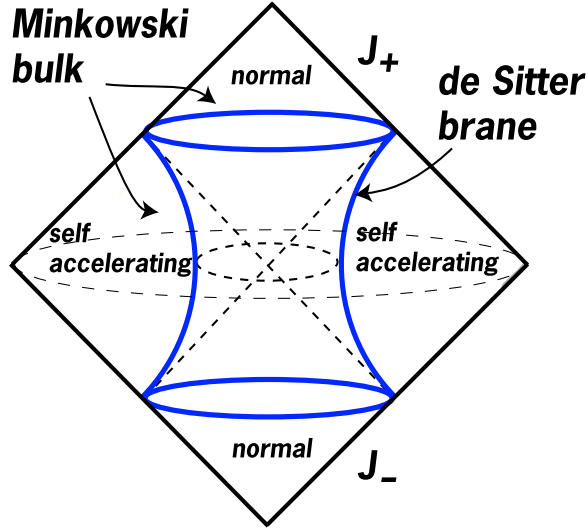, width=78mm, height=7cm}} 
\caption{\small Embedding of a general DGP brane in a
flat 5D bulk. The brane world volume is the hyperboloid in the
Minkowski bulk.} \label{fig:one}}
\end{figure}

The general DGP brane is a de-Sitter hyperboloid of curvature
\begin{equation}
H_\epsilon = \frac{M_5^3}{M_4^2} \left [
\sqrt{1 + \frac{M_4^2 \sigma}{6 M_5^6}} \; - \epsilon \right ]
\end{equation}
embedded in Minkowski spacetime, with the N-branch retaining the 
interior for its bulk, and the SA branch the exterior, see figure
\ref{fig:one}.

Although we can coordinatize the spacetime with explicitly
flat bulk Minkowski coordinates, it is more convenient to write the
cosmological DGP braneworld in brane-based coordinates, which make
the de-Sitter nature of the brane explicit:
\begin{equation}
\label{bgmetric}
ds^2= a^2(y) {\hat\gamma}_{ab} dx^adx^b =a^2(y)
\left( {\bar\gamma}_{\mu\nu}dx^\mu dx^\nu - dy^2\right) \, , 
\end{equation}
with ${\bar\gamma}_{\mu\nu}$ one of the standard de-Sitter metrics,
and $a(y) = e^{-\epsilon H_\epsilon |y|}$.

\section{DGP Specteroscopy\cite{CGKP}}

The first problem identified with the DGP model was the presence of 
ghosts. Initially, the DGP model appeared ghost free (unlike its
resolved cousin the GRS model) and therefore was promising as a
cosmological braneworld, however, while it is true that the normal
branch of DGP is ghost-free, the SA branch is not, although the full
picture is actually quite detailed and subtle. I will present a
summary of the salient features here, but see \cite{CGKP,DGPghost} for 
full detail.

In order to check the fluctuation spectrum, one perturbs the
background metric and the brane:
\begin{equation}
ds^2= a^2(y) \Bigl(\hat \gamma_{ab}+ a(y)^{-3/2}
h_{ab}(x,y)\Bigr)dx^adx^b \, , \qquad y = F(x^\mu) \label{perts}
\end{equation}
where the metric perturbations are obviously gauge dependent. 
It is convenient to work in the Gaussian Normal gauge ($h_{ay}=0$), 
and to further decompose the perturbation into its irreducible
components:
\begin{equation} 
h_{\mu\nu} = h^{\tt TT}_{\mu\nu} + D_\mu A_\nu + D_\nu A_\mu +
D_\mu D_\nu \phi - \frac14 \bar \gamma_{\mu\nu} D^2 \phi +
\frac{h}{4} \bar \gamma_{\mu\nu} \, , \label{decomps} 
\end{equation}
in which $h^{\tt TT}_{\mu\nu}$ is a transverse tracefree spin two perturbation,
$A_\mu$ a Lorentz gauge vector, and $\phi$ and $h = h^\mu_\mu$ are two
scalars. Note: this decomposition is only unique is different irreducible
representations have different masses. As we will see, if masses are 
degenerate, some mixing can occur between such components. 

Clearly, there is still further gauge freedom, and we can gauge away
the vector component, and shift the brane to the origin leaving:
\begin{equation} 
h_{\mu\nu} = h^{\tt TT}_{\mu\nu} + \left ( {\cal O}_{\mu\nu} - \quarter
{\cal O}^\lambda{}_\lambda {\bar \gamma}_{\mu\nu} \right ) \phi +
\frac{2a^{1/2}}{\epsilon H} {\cal O}_{\mu\nu} F + \quarter h
{\bar\gamma}_{\mu\nu} \, . 
\end{equation}
Here, ${\cal O}_{\mu\nu} = D_\mu D_\nu - H^2 {\bar\gamma}_{\mu\nu}$, and
$F$ is the function representing the fluttering of the brane. 

We can now solve for these perturbations, separating variables into
mass eigenmodes, and transverse eigenvalue equations in the $y$ variable.
For the TT mode we get:
\begin{equation}
h_{\mu\nu} (x^\mu, y) \sim u_m(y) \, \chi_{\mu\nu}^{(m)} (x) \, , \qquad 
[D^2 + 2H^2 ] \chi_{\mu\nu}^{(m)} = - m^2 \chi_{\mu\nu}^{(m)}
\end{equation}
with the eigenvalue equation
\begin{equation} 
u_m'' + \left [ m^2 - \frac{9H^2}{4} + \left ( \frac{M_4^2}{M_5^3} \, m^2
+3\epsilon H \right ) \delta (y) \right ] u_m = 0
\end{equation} 
which gives a discrete mode localized to the delta-function, and a continuum
gapped by $m>3H/2$. The discrete mode solution is:
\begin{equation} 
u_m(y) = \alpha_m e^{-\lambda_m y}
\end{equation} 
where
\begin{equation} 
\alpha_m = \frac{1}{M_4} \left [ \frac{3M_4^2 H - 2M_5^3 (1-\epsilon)}
{3M_4^2 H - 2M^3_5 \epsilon} \right ] ^{1/2} \, ,\;\;
\lambda_m = \sqrt{\frac{9H^2}{4} - m_d^2} \,,\;\;
\end{equation}
and
\begin{equation}
m_d^2 = (1-\epsilon) \frac{M_5^3}{M_4^2} \left [ 3H - \frac{2M_5^3}{M_4^2}
\right ]
\end{equation} 
is the mass of the discrete mode. Note that this is zero for the normal
branch, but for the SA branch:
\begin{equation} 
m_d^2 = \frac{2M_5^3}{M_4^2} \left [ 3H - \frac{2M_5^3}{M_4^2}
\right ] \rightarrow \begin{cases} 0<m_d^2 < 2H^2  & {\rm for}\ \sigma>0; \cr
m_d^2 > 2H^2  & {\rm for}\ \sigma<0.\end{cases}
\end{equation} 
Thus, for the SA branch, the spin two mode can have a mass which lies in the
`forbidden range' \cite{higuchi}, and therefore the positive tension
SA brane has a zero helicity spin two ghost.

Computing the scalar mode for the SA branch, $h_{\mu\nu}^{(\phi)}(x, y)=
W(y) {\cal O}_{\mu\nu}\hat \phi(x) $, shows that this has a mass
of $m^2 = 2H^2$, and obeys a transverse eigenvalue equation:
\begin{equation}
W^{\prime\prime}(y)-\frac{H^2}{4}W(y)=0 \, . \label{waveq}
\end{equation}
The boundary condition at the brane enforces a relation between
${\hat\phi}$ and $F$:
\begin{equation}
\left ( W'(0) +  \Bigl(\frac32\epsilon H+
\frac{M_4^2H^2}{M_5^3} \Bigr) W(0) \right ) {\hat \phi}
= 2 \Bigl(1 +
\epsilon \frac{H M_4^2}{M_5^3} \Bigr) F \, .
\label{Wbc} 
\end{equation}
which apparently ties the brane motion to the scalar mode on the SA
branch:
\begin{equation}
h_{\mu\nu}^{(\phi)} =-\frac{2}{H}  \left[\frac{M_5^3-
M_4^2H}{2M_5^3-M_4^2H}\right] e^{-Hy/2}
{\cal O}_{\mu\nu} F \, = - \alpha a^{-1/2} {\cal O}_{\mu\nu} F \,.
\label{phiF} 
\end{equation}
Computing the kinetic term for this scalar shows that if the brane
has negative tension, then this mode is a ghost -- as might have been
expected. 

To sum up, the full perturbation is: 
\begin{eqnarray}
\label{homsoln} h_{\mu\nu}(x, y)= &&
\alpha_{m_d}e^{-\lambda_{m_d}y}\chi_{\mu\nu}^{(m_d)}(x)
+\int_{\frac{3H}{2}}^\infty dm~u_m(y) \chi_{\mu\nu}^{(m)}(x) \nonumber\\
&+& \frac{(1-\epsilon)}{H} \left \{ a^{1/2} {\cal O}_{\mu\nu} F - \left
[ \frac{M_5^3-M_4^2H}{2M_5^3-M_4^2H} \right] a^{-1/2} {\cal
O}_{\mu\nu} F \right \} \, . 
\end{eqnarray}
For positive tension there is a ghost in the spin 2 sector, and for
negative tension the scalar is a ghost, but what happens for the pure
SA universe? If $\sigma=0$, the mass of the spin 2 mode becomes
$m_d^2 = 2H^2$, and the negative norm ghost would, if this were simple
massive gravity, become a gauge state due to an additional
`accidental' symmetry \cite{desernep}:
\begin{equation}
\chi_{\mu\nu} \to \chi_{\mu\nu} + {\cal O}_{\mu\nu} \vartheta, 
\;, \qquad (D^2 - 4H^2 ) \vartheta = 0\;.
\end{equation}
Further, we have a scalar mode, also on the borderline of becoming a ghost,
in which the normalization constant, $\alpha$, in (\ref{phiF}) 
has a pole at $\sigma=0$. In the absence of any other information, we
would conclude that this requires $F=0$, and thus neither mode would be
physical at $\sigma=0$.
However, we are not dealing with massive gravity, but DGP-brane gravity,
and these modes are both part of the 5D graviton, and therefore cannot
be considered separately. Moreover, the scalar
mode has the same form as the Deser-Nepomechie symmetry, thus we must be
careful in taking the $\sigma\to0$ limit, in order that we correctly
disentangle this symmetry from the physical brane motion. 

To do this, we rewrite our perturbation slightly:
\begin{equation} 
\alpha_{m_d}\chi_{\mu\nu}^{(m_d)}(x) =  {\cal H}_{\mu\nu}(x) -  \alpha
\, {\cal O}_{\mu\nu} F \, . \label{shiftchi} 
\end{equation} 
(where $\alpha$ is defined in (\ref{phiF})).
Substituting this in (\ref{homsoln}), and carefully taking the 
$\sigma \rightarrow 0$ limit (noting
that $\alpha \propto 1/\sigma$, and $\lambda_{m_d} = H/2 + O(\sigma)$)
\begin{equation} 
\label{notensionsoln} 
h_{\mu\nu}(x, y) = e^{-\frac{H}{2}y}
\Bigl({\cal H}_{\mu\nu}(x) - y \, {\cal O}_{\mu\nu} F \Bigr) +
\int_{\frac{3H}{2}}^\infty dm~u_m(y) \chi_{\mu\nu}^{(m)}(x)
+ \frac{2}{H} e^{Hy/2} {\cal O}_{\mu\nu} F\, , 
\end{equation} 
where ${\cal H}_{\mu\nu}$ satisfies 
\begin{equation} 
(D^2 +4H^2) {\cal H}_{\mu\nu}=- H {\cal O}_{\mu\nu} F \, .
\label{Feqsgh}
\end{equation} 
Thus by taking the limit properly, the physical information that is the
brane motion is retained, but now it is explicitly coupling in to the zero
helicity component of the 4D graviton, with the `scalar' part of the 
perturbation absorbing the DN symmetry. We now have the correct number of
physical degrees of freedom, and the remaining mode (\ref{Feqsgh}) is
indeed a ghost\cite{DGPghost,CGKP}.

\section{Brane Hypertension and Energy Problems\cite{GKMP}}

Recall that the most general cosmological DGP brane satisfied the
pseudo-Friedmann equation (\ref{frweq}). The general solution to
this equation with nonzero $\mu$ and $\kappa$ must be found numerically,
however, we can get a qualitative picture by plotting a phase plane
or $H$ and $1/R$, which allows us to draw some surprising conclusions
about the general SA brane solutions.

\begin{figure}
{\centerline{\epsfig{file=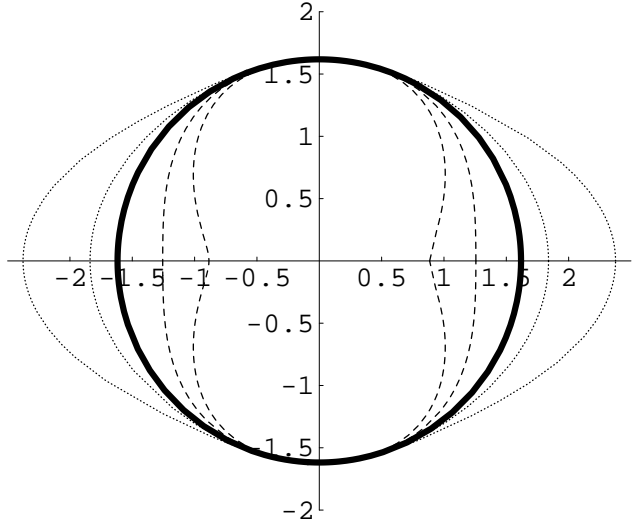, height=7cm}} 
\caption{\small Plot of the phase place of the general DGP brane in a
bulk with a black hole. The solid line is the pure accelerating universe
with no bulk black hole. The dotted lines are the DGP brane solution with
a {\it negative} mass bulk black hole, and the dashed lines those with a
{\it positive} mass bulk black hole.} \label{fig:phase}}
\end{figure}

First, note that since the SA brane is an hyperboloid in Minkowski
spacetime, it is natural to use global coordinates in which $\kappa=1$.
Next, for clarity, we absorb the various parameters in our variables by
defining
\begin{equation}
E = \frac{\sigma M_4^2}{24M_5^6} 
\;,\qquad {\hat\mu} = \frac{4M_5^6\mu}{M_4^4}
\;,\qquad X = \frac{M_4^2}{2M_5^3R}
\;,\qquad Y = \frac{M_4^2 H}{2M_5^3}
\end{equation}
thus squaring (\ref{frweq}) we get
\begin{equation}
\left ( X^2 + Y^2 \right )^2 - (1+2E) \left ( X^2 + Y^2 \right)
+ E^2 + {\hat\mu} X^4 = 0
\label{phaseeq}
\end{equation}
which gives the phase plot shown in figure \ref{fig:phase}.
It is easy to see that in the absence of a bulk black hole, the
trajectory is a circle in the $(X,Y)$ plane, whose radius is fixed by
the brane tension. We also see that for very small black hole masses,
the cosmology is very slightly perturbed, with the SA brane being
{\it repelled} by positive mass black holes, and {\it attracted} by negative
mass black holes. Also, notice that for large (positive) black hole mass, the
trajectory becomes pathological -- there is a turning point in $X$ without
a corresponding zero in $Y$. This is because this phase plane does not
represent an autonomous dynamical system, but rather an aid to understanding
general features of the solution. 

First of all consider negative mass bulk black holes. How can a negative
mass arise? Since the SA brane keeps the exterior of the hyperboloid
(see figure \ref{fig:one}), we can replace the flat spacetime with a
negative mass Schwarzschild spacetime without introducing any singularity
since ``$r=0$'' is not part of the spacetime. Thus while not allowed on
the normal branch, negative mass black holes are legitimate on the SA branch,
indeed, the SA brane is attracted towards such sources. Thus, as 
a 5D theory, the SA DGP branes admit solutions for which the
5D energy can be arbitrarily negative. Having identified
new negative energy configurations, it is natural to think that
these will be excited in both classical and quantum processes. While
finding explicit solutions which demonstrate the appearance of these
negative energy states in various dynamical processes is difficult,
it remains reasonable to assume that the theory should be
unstable.

Suppose however we ignore this problem, and somehow demand that there 
should be no negative mass black hole solutions even though they 
are physically sensible: this still leaves the positive mass black holes. 
Here we can see trajectories in figure \ref{fig:phase}
which are singular, in the sense that the scale factor has a minimum
(i.e.\ $X$ is bounded above) without a corresponding zero in the 
Hubble parameter (i.e.\ $Y$). By solving (\ref{phaseeq}) as a 
quadratic in $Y^2$:
\begin{equation}
Y^2 = E + \half - X^2 \pm \sqrt{E + \quarter - {\hat\mu}X^4}
\end{equation}
it is easy to see that this occurs if
${\hat\mu}X^4 = E+\quarter$, and $
{\hat\mu} \left ( E^2 + E + \quarter \right ) > E + \quarter $.
Clearly at such a point both $X$ and $Y$ are finite, but
differentiating (\ref{phaseeq}), and using ${\dot X}\propto -XY$ gives
\begin{equation}
{\dot Y} \propto - X^2
- \frac{2{\hat\mu} X^4 }{\left [ 2(X^2+Y^2) - (1+2E) \right ]}
\end{equation}
which can be seen to diverge at this singular point. However, if $\dot Y$
diverges, then ${\ddot R}/R$ must diverge, hence we have a true physical
curvature singularity. Since $H^2$ is finite, we see that the universe
retains finite energy, but its pressure becomes infinite at this crash.
This is reminiscent of the pressure singularities one obtains in finding 
exact static brane Tolman-Oppenheimer-Volkov solutions \cite{CGKM},
however, in that case it was due to the brane touching the bulk
event horizon. In this case, the critical bulk radius at which the 
brane ``crashes'' is 
\begin{equation}
R_c = \mu^{1/4} \left [ {\frac{M_5^6}{M_4^4} + \frac{\sigma}{6M_4^2}}
\right ] ^{-1/4}
\end{equation}
which is typically well outside the event horizon $R_h=\sqrt{\mu}$\cite{GKMP}.

To sum up, once we include the fully general bulk solution for DGP
brane cosmologies, the SA branch becomes catastrophically unstable, 
with solutions whose energy is unbounded below, and positive mass
bulk black holes with cosmologically crashing branes. It would be
tempting to postulate some sort of superselection rule for SA branes,
which fixed the mass of any bulk black hole to zero to avoid such
problems, however, if such a solution is to work, then we must not
be able to access nonzero mass bulk black holes via
perturbations of the SA brane. Unfortunately, it can be shown
that the bulk black hole corresponds to the scalar radion degree
of freedom in the perturbation theory of the brane \cite{GKMP},
and hence if we allow an SA cosmological DGP brane, we must also
allow the cosmological brane in the presence of the bulk black hole,
and therefore all of these pathologies.

\section{Hubble Bubble...Trouble with Instantons\cite{GKMP}}

\begin{figure}
{\centerline{\epsfig{file=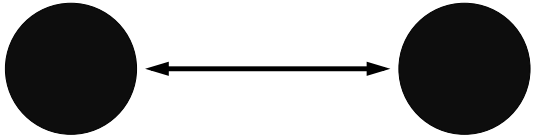, width=7cm}} 
\caption{\small Nucleation of a wormhole in the 5D vacuum by an 
SA brane connecting two different regions of Minkowski spacetime.} 
\label{fig:whole}}
\end{figure}

Finally, consider non-perturbative effects in the DGP model. The
SA brane, which is a hyperboloid plus exterior in Minkowski spacetime,
becomes a sphere plus exterior when rotated to Euclidean signature.
This instanton represents a possible decay of the 5D vacuum in the
presence of SA DGP branes. It is analogous to the `bubble of 
nothing'\cite{witten},
in that once the brane has been nucleated, it eats up the spacetime,
however, unlike the bubble of nothing it actually contains
another copy of the exterior spacetime on the other side of the brane.
In fact, if we relax the assumption of $\mathbb{Z}_2$ symmetry, about
the brane, then we could imagine nucleating an SA brane
which connects two different regions of space - a {\it wormhole}
(see figure \ref{fig:whole}). Such processes, resulting the the decay
of the vacuum are potentially disturbing, however, the crucial quantity
that determines how dangerous they are is their nucleation probability.
For this, we need to compute the Euclidean action:
\begin{equation}
I = S_E({\textrm{bubble}}) - S_E({\textrm{background}})
\end{equation}
where
\begin{equation}
S_E=-2M_5^3\int_\textrm{bulk} \sqrt{g} R
- 4M_5^3 \int_{\textrm{brane} + \infty} \sqrt{\gamma}  K
+ \int_\textrm{brane} \sqrt{-\gamma} \left [ 
- M_4^2 \mathcal{R} + \sigma \right]
\label{Eaction}
\end{equation}
Note that in the computation of the Gibbons-Hawking term, we have to 
also include the boundary at infinity, which is required for finiteness
of the Euclidean action after background subtraction. 
Clearly, the bulk (and therefore the boundary at infinity) is identical,
and Ricci flat, for both the SA bubble and the background, and thus the 
computation of the action reduces to the computation of the brane part
of (\ref{Eaction}). From the equations of motion of the brane, we have
\begin{equation}
\int_{\textrm{brane}}\sqrt{\gamma} 
\left ( 4M_5^3K+M_4^2 \mathcal{R}- \sigma \right )
= \int d\Omega_3\int^{\frac{\pi}{2H_-}}_{-\frac{\pi}{2H_-}}d\tau_E \, 
\frac{\cos(H_-\tau_E)^3}{3H_-^3} \left (M_4^2 \mathcal{R} + \sigma \right )
\end{equation}
and thus inputting $\mathcal{R} = 12 H_-^2$ for the Euclidean de Sitter
bubble, we get:
\begin{equation}
I = -\frac{8\pi^2}{9H_-^4}\left(12M_4^2H_-^2+\sigma\right)
\end{equation}
Note that the sign of this expression is {\it negative} for 
positive tension, $\sigma>0$. Therefore the tunneling amplitude
\begin{equation}
\mathcal{P} \propto e^{-I/\hbar} \, , \label{prob}
\end{equation}
would appear to be greater than one! What this means is that the
saddle point approximation used in the derivation of (\ref{prob})
is not valid, which indicates that tunneling is not suppressed 
and so there is large mixing
between empty 5D spacetime and that containing a
SA brane. By extension, one can infer that there is a large mixing
between all of the 5D spaces containing any number of
SA branes, and that empty five-dimensional
Minkowski space does not provide a good description of the quantum
vacuum of the five-dimensional theory. 

\section{Summary}

To summarize: the DGP model has been used widely
as a means of producing late time acceleration in the Universe. 
The key to providing this acceleration is choosing a particular 
{\it self-accelerating} `branch' of the brane cosmological solutions 
which accelerates even in the absence of matter. Unfortunately,
this branch has many physical problems:

\noindent $\bullet$ Strike 1: Ghosts are present in the perturbation
theory of SA branes. In most cases, the presence of a ghost would
immediately disqualify any theory or model from having physical
significance, although for DGP differing views have emerged as to the
severity of this pathology\cite{copout}, in particular, that strong
coupling might somehow save the day. It is rather unsatisfying however,
to have to appeal to strong coupling simply to have a single electron 
in your universe. 

\noindent $\bullet$ Strike 2: The SA branch of cosmological solutions
has pressure catastrophies, and its energy is unbounded below.
Thus, even if one simply regards the classical solution set of the DGP
branes, and ignores quantum consistency, there are severe problems. In
this case, strong coupling cannot save the day, as this problem arises
in the background solution.

\noindent $\bullet$ Strike 3: The SA branch destabilises the 5D vacuum
by unsuppressed tunneling processes which can either join alternate
5D bulks, or create wormholes in the vacuum. This last flaw is perhaps
the most controversial, since Euclidean quantum gravity is an imperfect
theory at best. However, the fact that the SA brane solutions give
instantons which do {\it not} correspond to stable saddle points in the
Euclidean theory is very worrying.

Interestingly, Izumi et.\ al.\ \cite{Izu} attempted to construct
configurations describing tunneling from SA to N branch branes,
such solutions require embedded domain walls \cite{ribbon} separating 
different de-Sitter phases, and they argued that such walls may not
exist, and hence that the ghost may not be so serious an instability of 
the SA brane. Note that the arguments presented here are orthogonal
to this investigation, and nonperturbative instabilities do exist.

Thus, although the DGP model was promising as a method of modifying
gravity, it seems that in the process of creating a good description
of late time 4D cosmology, it has destroyed the 5D Universe in which it
is embedded! Any one of these problems is serious, all three are fatal. 
Finally, it is worth stressing that these problems arise {\it only} with the
SA branch of solutions, the normal branch of DGP is not pathological. It
is still possible that the DGP model can be used to produce late time
acceleration for example by having asymmetric branes\cite{asy}.
The lesson seems to be that if modifying gravity with branes, 
one must always choose a normal branch.

\section*{Acknowledgements}
I would like to thank my collaborators, Christos Charmousis,
Nemanja Kaloper, Rob Myers and Tony Padilla, and the organizers for 
inviting me to such a lovely city.

%

\end{document}